\documentclass{article}

\usepackage[T1]{fontenc}

\usepackage{authblk}
\usepackage[noend]{algpseudocode}
\usepackage[backend=biber,style=alphabetic]{biblatex}
\usepackage{prelude}

\DefineBibliographyStrings{english}{in = {}}
\bibliography{2nd_spectral-gap}

\newcommand\formalise[1]{\guillemotleft\textit{#1}\guillemotright}

\begin{document}


\title{A Note on the Second Spectral Gap Incompleteness Theorem}

\author{Toby Cubitt}
\affil{Department of Computer Science, University College London, UK}
\date{}

\maketitle

\begin{abstract}
  Pick a formal system. Any formal system. Whatever your favourite formal system is, as long as it's capable of reasoning about elementary arithmetic.\footnote{Peano arithmetic + first-order logic? NBG set theory? New Foundations? Surely not boring old ZFC! And don't think you can wiggle out of it by telling me you're a constructivist: by the end of this abstract, you constructivists won't have escaped either.}
  The First Spectral Gap Incompleteness Theorem of \cite{spectral-gap} proved that there exist Hamiltonians whose spectral gap is independent of that system; your formal system is incapable of proving that the Hamiltonian is gapped, and equally incapable of proving that it's gapless.

  In this note, I prove a Second Spectral Gap Incompleteness Theorem: I show how to explicitly construct, within the formal system, a concrete example of a Hamiltonian whose spectral gap is independent of that system.
  Just to be sure, I prove this result three times.
  Once with G\"odel's help.
  Once with Zermelo and Fraenkel's help.
  And finally, doing away with these high-powered friends, I give a simple, direct argument which reveals the self-referential structure at the heart of these results, by asking the Hamiltonian about its own spectral gap.
\end{abstract}

G\"odel's famous 1931 results~\cite{Godel}, nowadays known as his First and Second incompleteness theorems, taught us that we cannot expect to prove everything we would like to in mathematics.
G\"odel's first incompleteness theorem proves that there exist statements in mathematics that can neither be proven nor disproven.
His second incompleteness theorem extends this to statements we \emph{might actually care about}.
More specifically, the First Incompleteness Theorem states (technically this is the slightly stronger Rosser's Theorem~\cite{Rosser}):
\begin{theorem}[G\"odel's 1st Incompleteness Theorem -- informal]\hfill\linebreak
  \label{Godel1}
  Let $\mathcal{F}$ be a consistent formal system capable of reasoning about elementary arithmetic.
  Then there exists a statement $S\in\mathcal{F}$ such that neither $S$ nor $\neg S$ are provable in $\mathcal{F}$.
\end{theorem}
This shows that there necessarily exist mathematical statements that are beyond the reach of proof.

The Second Incompleteness Theorem states:
\begin{theorem}[G\"odel's 2nd Incompleteness Theorem -- informal]\hfill\linebreak
  \label{Godel2}
  Let $\mathcal{F}$ be a consistent formal system capable of reasoning about elementary arithmetic.
  (ZFC is a popular choice.)
  Then $\mathcal{F}$ cannot prove its own completeness: the statement $G=$\formalise{$\mathcal{F}$ is complete}, formalised within $\mathcal{F}$, cannot be proven in $\mathcal{F}$.
\end{theorem}
This proves that a mathematical statement of real significance is beyond the reach of proof.
Namely, the statement of the system's own completeness.
(I will use \formalise{\dots} to indicate the formalisation of a colloquial mathematical statement within a formal system, and leave writing these out in full as fun exercises for the reader.)

In \cite{spectral-gap}, we showed that the same thing affects\footnote{Infects?}
certain important questions in theoretical physics.
Specifically, in~\cite{spectral-gap} we proved that the spectral gap problem is undecidable.
Indeed, in a nod to Rice's theorem~\cite{Rice}, we proved undecidability of any property that distinguishes gapped Hamiltonians with product ground states, from gapless Hamiltonians with algebraically-decaying connected correlation functions.
\cite{spectral-gap} proved this for quantum systems in 2D or larger.
In \cite{spectral-gap_1d} we strengthened these result to include 1D systems.

More precisely, we considered translationally invariant, nearest-neighbour spin lattice Hamiltonians $H^{\Lambda(L)} = \sum_{(i,j)\in\mathcal{E}} h^{(i,j)}+\sum_{k\in\Lambda(L)} h_1^{(k)}$ on a lattice $\Lambda(L)$ of size $L$, with ``gapped'' and ``gapless'' defined as follows:

\begin{definition}[Gapped]\label{def:gapped}
  We say that a family of Hamiltonians $H^{\Lambda(L)}$, as described above, is \emph{gapped} if there is a constant $\gamma>0$ and a system size $L_0$ such that for all $L>L_0$, $\lambda_0(H^{\Lambda(L)})$ is non-degenerate and $\Delta(H^{\Lambda(L)})\geq\gamma$. In this case, we say that \emph{the spectral gap is at least $\gamma$}.
\end{definition}

\begin{definition}[Gapless]\label{def:gapless}
  We say that a family of Hamiltonians $H^{\Lambda(L)}$, as described above, is \emph{gapless} if there is a constant $c>0$ such that for all $\varepsilon>0$ there is an $L_0\in\mathbb{N}$ so that for all $L>L_0$ any point in $[\lambda_0(H^{\Lambda(L)}),\lambda_0(H^{\Lambda(L)})+c]$ is within distance $\varepsilon$ from $\spec H^{\Lambda(L)}$.
\end{definition}

\noindent Under these mathematically precise definitions of ``gapped'' and ``gapless'', we proved the following result for 2D (and higher) systems~\cite{spectral-gap}:

\begin{theorem}[2D Spectral Gap Undecidability]\label{thm:2d}
  For any given Turing Machine $M$ and any $n\in \N$, we can explicitly construct a fixed positive integer $d$ and a fixed set of Hermitian matrices $h_1(n) \in \cB(\C^d)$ and $h_{\text{row}}(n),h_{\text{col}}(n) \in \cB(\C^d\ox\C^d)$ with algebraic matrix elements, where $\matnorm{h_1(n)},\matnorm{h_{\text{col}}(n)},\matnorm{h_{\text{row}}(n)} \leq 1$, such that:
  \begin{enumerate}
  \item
    If $M$ halts on input $n$, then the associated family of nearest-neighbour, translationally invariant, 2D spin lattice Hamiltonians
    \begin{equation}\label{eq:2d}
      H^{\Lambda(L)}(n) = \sum_{i\in\text{lattice}} h_1^{(i)}(n)
                  + \sum_{(i,j) \in \text{Rows}} h_{\text{row}}^{(i,j)}(n)
                  + \sum_{(i,j) \in \text{Cols}} h_{\text{col}}^{(i,j)}(n)
    \end{equation}
    is gapped in the sense of \cref{def:gapped}, with gap $\gamma\ge 1$.
  \item
    If $M$ does not halt on input $n$, then $H^{\Lambda(L)}(n)$ is gapless in the sense of \cref{def:gapless}.
  \end{enumerate}
\end{theorem}

\noindent We subsequently strengthened this to the following result for 1D chains~\cite{spectral-gap_1d}:

\begin{theorem}[1D Spectral Gap Undecidability]\label{thm:1d}
  For any given Turing Machine $M$ and any $n\in\N$, we can explicitly construct a fixed positive integer $d$ and a fixed set of Hermitian matrices $h_1(n) \in \cB(\C^d)$ and $h(n) \in \cB(\C^d\ox\C^d)$ with algebraic matrix elements, where $\matnorm{h_1(n)},\matnorm{h(n)} \leq 1$, such that:
  \begin{enumerate}
  \item
    If $M$ halts on input $n$, then the associated family of nearest-neighbour, translationally invariant, 1D spin chain Hamiltonians
    \begin{equation}\label{eq:1d}
      H^{\Lambda(L)}(n) = \sum_{1\leq i\leq L} h_1^{(i)}(n) + \sum_{0\leq i \leq L-1} h^{(i,i+1)}(n)
    \end{equation}
    is gapless in the sense of \cref{def:gapped}.
  \item
    If $M$ does not halt on input $n$, then $H^L(n)$ is gapped in the sense of \cref{def:gapless}, with gap $\gamma\ge 1$.
  \end{enumerate}
\end{theorem}

Letting $M$ be a universal Turing Machine in the above results, this immediately implies (algorithmic) undecidability of the spectral gap problem, via the classic result by \textcite{Turing} that the Halting problem is undecidable.
As explained in~\cite{spectral-gap}, by standard arguments this result also implies a form of axiomatic independence \`a la G\"odel:

\begin{corollary}[1st spectral gap incompleteness theorem]\label{1st_spectral}
  Let $\mathcal{F}$ be a consistent formal system capable of reasoning about elementary arithmetic.
  Let $d\in\mathbb{N}$ be a sufficiently large constant (whose numerical value can in principle be specified explicitly).
  Then there exist translationally invariant nearest-neighbour Hamiltonians on a 2D lattice, or on a 1D chain, with local dimension $d$ and algebraic matrix entries, for which the presence or the absence of a spectral gap is unprovable in $\mathcal{F}$.
\end{corollary}

This is (sort of) a spectral-gap analogue of the usual statement of G\"odel's First Incompleteness Theorem:\footnote{
  Don't take this analogy -- or this note -- too seriously!
  There isn't a formal analogy in any rigorous sense between G\"odel's Incompleteness Theorems and the Spectral Gap Incompleteness Theorems.
  It's just an analogy in the loose, informal, colloquial sense.
  I.e.\ the usual English sense.
  Not some deep and meaningful mathematical sense.

  G\"odel's First Incompleteness Theorem is usually stated something along the lines of \cref{Godel1},\stepcounter{footnote}\footnotemark[\thefootnote] asserting existence of unprovable statements $S$.
  But, unlike the First Spectral Gap Incompleteness Theorem, the proof of G\"odel's First Incompleteness Theorem actually constructs an explicit example of such an $S$.
  It's just not a very interesting $S$.

  In case you're wondering, there's no formal or deep-and-meaningful analogy between consistency/inconsistency and gapped/gapless, either.
  \label{fn:analogy}}
\footnotetext[\thefootnote]{In any reference you should trust, it will be stated more precisely than here.\label{fn:Godel_ref}}
it proves existence of a spin lattice Hamiltonian with undecidable spectral gap.
But it isn't able to pinpoint any \emph{specific} Hamiltonian whose spectral gap is undecidable.
We noted in \cite{spectral-gap} that:

\begin{quote}
``[T]here are particular Hamiltonians within these families for which one can neither prove nor disprove the presence of a gap, or of any other undecidable property. Unfortunately, our methods cannot pinpoint these cases.''
\end{quote}

The purpose of this note is to point out that our methods \emph{can} in fact pinpoint concrete cases.
We'll prove a spectral-gap analogue\footnote{See \cref{fn:analogy}.} of G\"odel's Second Incompleteness Theorem: a concrete construction of a \emph{particular} many-body quantum Hamiltonian for which the presence or absence of a spectral gap cannot be proven.

I will assume\footnote{I.e.\ have already assumed.}
the reader understands (at least at an informal level) what a Turing Machine is, what the Halting Problem is and why it's undecidable, what a formal system is, what it means to ``formalise'' mathematical statements within the system, and the distinction between proof and truth.\footnote{Nothing too deep, then.}
If not, one or both of \cite{GEB} or \cite{Oberhoff} are required reading.
The former inspired everyone who's read it.
The latter was the inspiration for this note.

\section{Second Spectral Gap Incompleteness Theorem}

We deliberately formulate the Second Spectral Gap Incompleteness Theorem in a way that mirrors its hitherto more famous forebear:\footnote{``Mirror'' has no formal or rigorous meaning here. See \cref{fn:analogy} if you're tempted to look for one.}

\begin{theorem}[2nd Spectral Gap Incompleteness Theorem -- informal]
  Let $\mathcal{F}$ be your favourite formal system capable of reasoning about mathematics.
  (ZFC is a popular choice.)
  Assuming $\mathcal{F}$ is consistent,\footnote{If you're assuming your favourite formalisation of mathematics is inconsistent, then you really have no business using it to reason about mathematics.
    In any case, if you're assuming it's inconsistent, then I can prove anything I like with it.
    If you're assuming it's consistent but is able to prove its own inconsistency, then you must be a professional mathematical logician. I'll deal with you in a moment.}
  we can explicitly describe within $\mathcal{F}$ a specific, translationally invariant, nearest-neighbour, spin lattice Hamiltonian for which the presence of a spectral gap cannot be proven in~$\mathcal{F}$.
  Similarly, we can explicitly describe within $\mathcal{F}$ a specific, translationally invariant, nearest-neighbour, spin lattice Hamiltonian for which the absence of a spectral gap cannot be proven in~$\mathcal{F}$.
\end{theorem}

G\"odel's Second Incompleteness Theorem is usually stated something along the lines of \cref{Godel2} (but see \cref{fn:Godel_ref}), giving a concrete and interesting example of an unprovable statement. The Second Spectral Gap Incompleteness Theorem gives a concrete example of a Hamiltonian whose spectral gap is unprovable. (Note the deliberate omission here, then read \cref{fn:interesting}.)

\noindent More precisely, by the end of this note we will have proven:
\begin{theorem}[2nd Spectral Gap Incompleteness Theorem]\hfill\linebreak
  \label{thm:2nd_spectral}
  Let $\mathcal{F}$ be any consistent, recursive formal system capable of reasoning about mathematics (such as ZFC).
  Assuming consistency of $\mathcal{F}$, we can explicitly construct within $\mathcal{F}$ a fixed positive integer $d$ and a fixed set of Hermitian matrices $h_1\in\cB(\C^d)$ and $h_{\text{row}},h_{\text{col}},h \in \cB(C^d)$ with algebraic matrix elements, where $\matnorm{h_1},\matnorm{h_{\text{row}}},\matnorm{h_{\text{col}}},\matnorm{h} \leq 1$, describing an associated family of nearest-neighbour, translationally invariant, 2D spin lattice Hamiltonians \cref{eq:2d} or 1D spin chain Hamiltonians \cref{eq:1d} $H^{\Lambda(L)}$, such that:
  \begin{enumerate}
  \item
    The statement $G\vee C =$\formalise{($H^{\Lambda(L)}$ is gapped) OR ($H^{\Lambda(L)}$ is gapless)} with gapped and gapless defined in the sense of \cref{def:gapped,def:gapless}, formalised within $\mathcal{F}$, can be proven within $\mathcal{F}$.
  \item
    The statements $G=$\formalise{$H^{\Lambda(L)}$ is gapped} and $C=$\formalise{$H^{\Lambda(L)}$ is gapless}, formalised within $\mathcal{F}$, are independent of $\mathcal{F}$.
  \end{enumerate}
\end{theorem}

In fact, we'll see three ways to prove some form of axiomatic independence of the spectral gap.
The first proof is a straightforward consequence of the original undecidability of the spectral gap results of \cref{thm:2d,thm:1d}.
The second proof uses results by \cite{Yedidia-Aaronson} to make the first proof more concrete and compelling.
However, neither of these proofs is quite strong enough to prove \cref{thm:2nd_spectral}, unless you are willing to make an additional assumption that your favourite formulation of mathematics is not insane.
The main point of this note is to observe that there is an elegant, self-contained argument that gives a Second Spectral Gap Undecidability Theorem directly, and as a bonus it does this without needing any additional assumptions on the sanity of mathematics.

\section{First proof attempt\protect\footnote{Tastefully numbered ``2''.}}
The first proof makes use of none other than G\"odel's Second Incompleteness Theorem itself, to prove a spectral gap analogue.

\begin{proof}[of \cref{thm:2nd_spectral}]
Since $\mathcal{F}$ is recursive, we can
construct a Turing Machine $M$ that enumerates over all the provable statements in $\mathcal{F}$, and halts if it ever finds a contradiction (i.e.\ if it ever finds both a proof of $T$ and a proof of $\neg T$).
In pseudocode:
\begin{algorithmic}
  \Function{$M$}{}
    \State $S\gets\emptyset$
    \ForAll{$x\in$ all strings in lexicographic order}
      \If{$\exists T \in S$ such that $\neg T\in S$}
        \State halt
      \ElsIf{$x$ is a proof of $T$}
        \State $S\gets S \cup \{T\}$
      \EndIf
    \EndFor
  \EndFunction
\end{algorithmic}
Assuming $\mathcal{F}$ is consistent, it will never find such a contradiction, so $M$ will never halt.
But, by G\"odel's Second Incompleteness Theorem (\cref{Godel2}), there is no way of proving that $\mathcal{F}$ is consistent within $\mathcal{F}$, hence no way to prove that $M$ will never halt.
I.e.\ \formalise{$\mathcal{F}$ is consistent} is independent of $\mathcal{F}$ by \cref{Godel2}, hence \formalise{$M$ runs forever} is independent of $\mathcal{F}$.
Thus $M$ is a \emph{particular} Turing Machine which must run forever if the system $\mathcal{F}$ we're using to reason about mathematics is consistent, but which we can never \emph{prove} will run forever using our chosen system of mathematics.

This argument is standard and well-known (see e.g.\ \cite{Oberhoff}).
But the short step from this standard argument to the spectral gap problem is already provided by \cref{thm:2d,thm:1d}!
(And all the steps in that argument can be formalised within any $\mathcal{F}$ sufficient to reason about mathematics, such as ZFC.)

Applying \cref{thm:2d} to $M$, we obtain a description of a \emph{specific} spin lattice Hamiltonian\footnote{Or more precisely, a description of its local interactions.}
$H$ which must be gapless if the system $\mathcal{F}$ we're using to reason about mathematics is consistent, but which we can never \emph{prove} is gapless using our chosen system of mathematics.
I.e.\ \formalise{$H$ is gapless} is independent of $\mathcal{F}$.

Similarly, we can apply \cref{thm:1d} to $M$ (or the construction in \cite[Section~2]{spectral-gap}, which inverts the relationship between halting/non-halting and gapped/gapless) to obtain a description of a different specific spin lattice Hamiltonian $H$ which must be gapped if $\mathcal{F}$ is consistent, but for which \formalise{$H$ is gapped} cannot be proven in $\mathcal{F}$.
\end{proof}

However, there's a subtle but important caveat with this argument, which means it fails to prove \cref{thm:2nd_spectral} in full.\footnote{Did you believe it was a full proof of \cref{thm:2nd_spectral} just because it said ``proof of \cref{thm:2nd_spectral}'' at the beginning?
  Never trust what a mathematician says! Always read the proof, and compare what was actually proven to what was \emph{claimed} to be proven in the theorem statement.}
G\"odel's Second Incompleteness Theorem tells us that $\mathcal{F}$ cannot prove its own \emph{consistency}.
However, it's perfectly possible for a consistent formal system to prove its own \emph{in}consistency!
If you're not a professional mathematical logician, then your brain may just have exploded.
To put your brain back together again, you need to keep in mind the difference between ``proof'' and ``truth'', and the meaning of ``consistent''.
Or more usefully, the difference between the formalisation of a mathematical statement -- a string of symbols -- and what we interpret those symbols to \emph{mean}.

G\"odel's theorem says that a consistent formal system $\mathcal{F}$ cannot be used to derive the string of symbols \formalise{$\mathcal{F}$ is consistent}.
On the other hand, it says nothing about the string of symbols \formalise{$\mathcal{F}$ is inconsistent}.
However, just because $\mathcal{F}$ can be used to derive this string of symbols, doesn't necessarily mean $\mathcal{F}$ is \emph{actually} inconsistent!
Perhaps $\mathcal{F}$ just lies about itself.

An inconsistent theory is, by definition, one which can prove a contradiction; i.e.\ one which can prove both a statement $P$ and the negation $\neg P$ of the same statement.
Just because $\mathcal{F}$ can prove the statement $I$=\formalise{$\mathcal{F}$ is inconsistent}, that doesn't imply that $\mathcal{F}$ is able to prove a contradiction.
Indeed, we already know (by G\"odel's theorem) that $\mathcal{F}$ can't prove the statement $\neg I=$\formalise{$\mathcal{F}$ is consistent} even if it's consistent.
So it's entirely consistent for a consistent formal system to prove its own inconsistency!
``Proof'' does not equal ``truth'' in mathematical logic.

A formalisation of mathematics that tells big, fat lies like this about itself is certainly not a very useful formalisation of mathematics.
We could never trust anything it told us.
Or, if we're feeling more generous to $\mathcal{F}$, the formal system itself is fine (it's consistent); it's the \emph{meaning} we're trying to attach to the strings of symbols it produces that isn't valid.
We live in hope that ZFC and other popular formalisations of mathematics don't suffer from this flaw.
But we can never know for sure because\dots G\"odel's Second Incompleteness Theorem.

G\"odel encountered similar subtleties in the original proof of his incompleteness theorems.
To get around this, he used the age-old mathematician's trick: define the problem away.
The original formulation of the incompleteness theorems assume something slightly stronger than consistency: they assume the forrmal system $\mathcal{F}$ is ``$\omega$-consistent''.
$\omega$-consistency amounts to assuming that the system doesn't lie in this way.
A formalisation of mathematics which fails to be $\omega$-consistent (i.e.\ one which is allowed to lie about basic statements in arithmetic) is not all that much better than a formalisation that is inconsistent (one which can prove a contradiction).
So you don't lose all that much by making this stronger assumption: G\"odel's original versions of the incompleteness theorems still imply that any formalisation of mathematics that a mathematician would ever want to use in practice, necessarily cannot prove all true statements about arithmetic.
Rosser~\cite{Rosser} later found a neat trick (now imaginatively called ``Rosser's trick'') to get around this technicality, and prove the incompleteness theorems assuming only consistency, rather than the stronger $\omega$-consistency.

However, our first attempt at a proof of the Second Spectral Gap Incompleteness Theorem uses a G\"odel-like statement, rather than a Rosser-like statement.
So although we can successfully argue that we can construct a particular gapped Hamiltonian which can never be proven to be gapped, it's possible that we might be able to prove that this particular Hamiltonian is \emph{gapless} (even though it isn't).
And, mutatis mutandis for the gapless Hamiltonian.
Unless we add an additional assumption that our formalisation of mathematics doesn't tell lies.
But then we have proven a weaker result than the one claimed in \cref{thm:2nd_spectral}.
(Not to mention one that's even harder to wrap one's head around!)

\section{Second proof attempt}
In principle, if you pick your favourite formal system $\mathcal{F}$ sufficiently powerful to reason about elementary arithmetic -- say ZFC -- the above argument will give you back a concrete description of a spin lattice Hamiltonian $H$ whose spectral gap status is independent of that formal system.
The procedure for explicitly constructing $H$ from $M$ is essentially the content of \cite{spectral-gap} (plus \cite{spectral-gap_1d} for the 1D case).
But one still needs to construct $M$ from $\mathcal{F}$.
It's clear enough that one \emph{could} in principle construct $M$ from any given $\mathcal{F}$.
But we haven't done so.
You might quibble that this falls somewhat short of our goals.
Strictly speaking, it doesn't completely specify a concrete procedure for constructing $H$; it shows how one \emph{could} fill in the gaps and come up with a complete procedure, were one so inclined.

Luckily, other people were so inclined.
E.g.\ \cite{Yedidia-Aaronson} gave an explicit construction of a concrete, 7910-state Turing Machine that implements $M$ for ZFC.
Plugging their Turing Machine into \cite{spectral-gap,spectral-gap_1d} then gives a complete procedure for constructing a specific gapped Hamiltonian $H$ which cannot be proven to be gapped in ZFC.

In the fine tradition of mathematics papers since time immemorial, I leave the now trivial task of writing out the matrix entries of the local interactions of $H$ as an exercise for the reader.

However, this argument (and also the \cite{Yedidia-Aaronson} result itself) still suffers from the issue discussed above: that ZFC might tell lies.
I.e.\ it's logically possible that ZFC can prove that $M$ halts, even though it doesn't, and hence prove that the Hamiltonian $H$ is gapped, even though it isn't.
Now, if ZFC -- the most commonly used foundation of modern mathematics -- is really able to lie about statements in arithmetic like this, then we have far bigger problems than the spectral gap one.
So in an informal way, by formalising and constructing everything explicitly for ZFC in particular, we've managed to pinpoint specific Hamiltonians whose spectral gap status is independent of ZFC unless most of modern mathematics comes crashing to the ground.
Nonetheless, whilst it makes things more explicit for the most popular\footnote{Or perhaps not the most popular.
  Perhaps ZFC is like email: widely used, but unpopular.}
foundation of mathematics, this second proof is technically no stronger than our first attempt,

\section{Third proof lucky}
The above proof(s) ultimately rely on G\"odel's Second Incompleteness Theorem, which tell us that consistency of any reasonable formalisation $\mathcal{F}$ of mathematics is unprovable within that system, to imply that non-halting of a specific $M$ is independent of $\mathcal{F}$.
It's a perfectly serviceable proof.
But as proofs go, it's slightly unsatisfying.
As well as proving a somewhat weaker result than claimed, the self-referential structure that lies at the heart of all incompleteness proofs~\cite{GEB} isn't evident.
It's there, but only indirectly, through bootstrapping off G\"odel's original Incompleteness Theorems (whose proofs of course \emph{do} have this self-referential structure).

Moreover, G\"odel's Second Incompleteness Theorem follows from the First Incompleteness Theorem.
But we already proved a First Spectral Gap Incompleteness Theorem in \cref{1st_spectral}.
Shouldn't there be some simpler, more elegant, direct argument?
One that directly lifts the First Spectral Gap Incompleteness Theorem to the Second Spectral Gap Incompleteness Theorem, instead of resorting to G\"odel's results?
And which also fixes up the proof to avoid the need for any additional consistency assumptions, at the same time?

Indeed there is!
A delightful recent essay by \textcite{Oberhoff} explains how G\"odel's First and Second Incompleteness Theorems can be proven directly from the connection between algorithms and formal systems that we just saw.
In a similar spirit, both the First and Second Spectral Gap Incompleteness Theorems can be proven directly from the connection between spectral gaps and formal systems established by \cite{spectral-gap_short,spectral-gap}.

This isn't a direct analogue of G\"odel's Second Incompleteness Theorem; there's no formal or rigorous analogy between the Incompleteness Theorems and the Spectral Gap Theorems.\footnote{If you don't like the name, feel free to call it the ``One-and-a-half'th Spectral Gap Incompleteness Theorem''. Or call it Fred.}
But the idea behind this third proof is somewhat reminiscent of how one lifts the First Incompleteness Theorem to the Second Incompleteness Theorem, by formalising the proof of the First Incompleteness Theorem within the formal system itself.
So I'm sticking to my name for it: the Second Spectral gap Incompleteness Theorem.

The proof here has to go via Turing Machines as an intermediate step, only because the results of \cite{spectral-gap,spectral-gap_1d} are phrased in terms of Turing Machines and not formal systems.
If you have the patience and energy (I don't!), you can redo the construction of \cite[Section~4]{spectral-gap} to directly encode inference within a given formal system into the Hamiltonian.
(Rather than first encoding inference into a Turing Machine, and then encoding that Turing Machine in turn into a Hamiltonian.)
By \cite{Turing}'s results, we know these are all the same thing anyway.

I'll give the argument for the 1D case, as this is the stronger result.
(Because you can always get a 2D system by placing lots of 1D systems next to each other.)
The argument using the original the 2D construction is almost identical, just flipping the construction in the obvious place since the relationship between gapped/gapless and halting/non-halting in the original 2D result of~\cite{spectral-gap} is inverted compared to the 1D result of~\cite{spectral-gap_1d}.

For convenience, let's give names to the statements that $H$ is gapped/gapless, formalised within $\mathcal{F}$. As before, we'll denote formalisation within $\mathcal{F}$ by \formalise{\dots}.
Let \mbox{$G(H)=$\formalise{$H$ is gapped}} ($G$ for ``gapped'') and \mbox{$C(H)=$\formalise{$H$ is gapless}} ($C$ for ``continuous spectrum'').
Note that $G(H)\neq \neg C(H)$, because our strong definitions of gapped and gapless (\cref{def:gapped,def:gapless}) are not negations of each other; they deliberately exclude ambiguous intermediate cases, such as Hamiltonians with degenerate ground states.
So we'll want to separately prove independence of both $G(H)$ and $C(H)$ from $\mathcal{F}$.
It is true that $C(H) \Rightarrow \neg G(H)$ and $G(H) \Rightarrow \neg C(H)$, so a consistent theory cannot prove both $G(H)$ and $C(H)$.
But, for any given $H$, that would still leave the possibility that both are false for the $H$ we construct, and that falsehood of both $C(H)$ and $G(H)$ is provable in $\mathcal{F}$.
We want to rule out this possibility, too, because we're in mathematical logic mode and want to take care of all possibilities, however mind-bendingly crazy mathematics would have to be if it actually behaved that way.

\begin{proof}[of \cref{thm:2nd_spectral}]
  Let $\mathcal{F}$ be any consistent, recursive formal system that is sufficiently powerful to reason about elementary arithmetic.
  Let $M$ be a Turing Machine that takes as input a description of a translationally-invariant spin-lattice Hamiltonian $H$, and enumerates over all possible proofs in $\mathcal{F}$ until it finds a proof of $G(H)$.
  In pseudocode:
  \begin{algorithmic}
    \Function{$M$}{$H$}
      \For{$x \in$ all strings in lexicographic order}
        \If{$x$ proves $G(H)$} halt
        \EndIf
      \EndFor
    \EndFunction
  \end{algorithmic}

  Now, let $H_M$ be the Hamiltonian obtained by applying \cref{thm:1d} to $M$ and setting $n$ in \cref{thm:1d} to the description of $H_M$.
  I claim that $G(H_M)$ is independent of $\mathcal{F}$.\footnote{See where the self-referentiality is coming in?}

  Assume for contradiction that \mbox{$G(H_M)=$\formalise{$H_M$ is gapped}} is provable in $\mathcal{F}$.
  Then $M$ will eventually find a proof of this, and halt.
  By \cref{thm:1d}, if $M$ halts then $H_M$ is gapless.
  So far, we've just shown that $\mathcal{F}$ necessarily tells lies about some spectral gaps.
  Just because it lies, doesn't mean that it's inconsistent.

  However, following~\cite{Oberhoff}, once you strip away the window-dressing, the entire proof of \cref{thm:1d} in \cite{spectral-gap,spectral-gap_1d} involves nothing more than elementary arithmetic and first-order logic,\footnote{The only reason we followed the standard mathematical convention of not reducing the whole proof down to Peano's axioms and first order logic was to keep the already-large page count of \cite{spectral-gap} down a bit.
    Quite a bit.
    In the British English sense of ``quite''.}
  hence can be formalised within $\mathcal{F}$.
  Since \cref{thm:1d} proves that $H_M$ is gapless iff $M$ runs forever, this, together with the sequence of computational steps that led to $M$ halting, constitutes a proof within $\mathcal{F}$ of $C(H_M)$.
  But we already saw that $C(H_M) \Rightarrow \neg G(H_M)$, so this contradicts consistency of $\mathcal{F}$.

  Now assume for contradiction that \mbox{$\neg G(H_M)=$\formalise{$H_M$ is not gapped}} is provable in $\mathcal{F}$.
  Since we're assuming $\mathcal{F}$ is consistent, this implies $M$ will never find a proof of $G(H_M)$, so will never halt, \cref{thm:1d} tells us that if $M$ doesn't halt, then $H_M$ is gapped, this gives us a proof formalisable within $\mathcal{F}$ of $G(H_M)$, and we again have a contradiction.

  Exactly the same argument goes through, mutatis mutandis, for $C(H_M)$.
  Thus, if $\mathcal{F}$ is consistent,\footnote{Recall that if $\mathcal{F}$ is \emph{in}consistent, then by well-known arguments we can prove literally every possible mathematical statement.
    So nothing at all can be independent of an inconsistent $\mathcal{F}$, let alone statements about spectral gaps.
  So as usual, we can only prove meaningful spectral gap independence results for consistent systems.
And we have to assume consistency rather than prove it, because \dots G\"odel's Second Incompleteness Theorem.}
  then both $G(H_M)$ and $C_H(M)$ are independent of $\mathcal{F}$, as claimed.
\end{proof}

If you haven't seen this type of thing before, you may be worried that there's an issue here.
The description of $H_M$ depends on $n$, which in turn depends on the description of $H_M$. Doesn't this mean $H_M$ is not well-defined?
On the other hand, if you know too much computability theory, you may think we need to appeal here to Kleene's second recursion theorem~\cite{Kleene}.

But this is no more of an issue than constructing a Turing Machine $B$ that, when fed the description of a Turing Machine $A$ as input, searches for proofs that the Turing Machine $A$ halts on input $A$ (as e.g.\ in \cite[Lemma~1]{Oberhoff}).
This is obviously fine, as $B$ is given the string $A$ as input, and is free to interpret it both as a description of a Turing Machine $A$, and as a string $A$ to be input to that Turing Machine.
If this still worries you for some reason, we could always have $B$ duplicate its input string $A$ as a first step, and use one copy as the description of a Turing Machine, the other as the input string to that machine.

The Hamiltonian case is very similar.
$M$ should take as input a string describing $H(n)$ with $n$ a free parameter, or ``placeholder'', and have $M$ interpret this input string both as a description of $H(n)$ and as the value of $n$.
By now you're probably not worried.
But just in case you're of a particularly nervous disposition, we could always have $M$ duplicate its input string as a first step, and use one copy as the description of $H(n)$, the other as the value of $n$.

This direct third proof is barely longer than the first one (upon which the second one builds), even though we've proven it directly as a stand-alone result, rather than reaching for the sledgehammer of G\"odel's Second Incompleteness Theorem.\footnote{Indeed, we've independently proven a theorem that's \emph{better}\stepcounter{footnote}\footnotemark[\thefootnote] than G\"odel's original:
we've proven that there are specific statements of interest to physicists -- about spectral gaps, no less! -- that are independent of whatever formal system floats your boat.}
\footnotetext[\thefootnote]{Unless you consider consistency of ZFC and statements about the logical structure of all mathematics to be more interesting than spectral gaps.
  But that highly unlikely possibility scarcely seems worth addressing.\label{fn:interesting}}
It also makes the self-referential structure of the proof evident: essentially, the proof constructs a Hamiltonian which can answer questions about spectral gaps through its spectral properties, then asks this Hamiltonian about its own spectral gap.

\printbibliography

\end{document}